\documentclass{article}
\usepackage{epsfig}

\date{\today} 

\begin{document}

\begin{center}
{\Large\bf NEW HAIRY BLACK HOLE SOLUTIONS \\ WITH A DILATON POTENTIAL}
\vspace{0.5cm}
\\
{\bf Eugen Radu$^{1}$ }
{\bf and D. H. Tchrakian $^{1,2}$ }

\vspace*{0.2cm}
{\it $^1$ Department of
Mathematical Physics, National University of Ireland Maynooth, Maynooth, Ireland}

{\it $^{2}$School of Theoretical Physics -- DIAS, 10 Burlington
Road, Dublin 4, Ireland}
\vspace{0.5cm}
\end{center}
\begin{abstract}
We consider black hole solutions
with a dilaton field possessing  a nontrivial
potential approaching a constant negative
value at infinity.
The asymptotic behaviour of the dilaton field is assumed to be
slower than that of a localized
distribution of matter.
A nonabelian SU(2) gauge field is also included in the total action.
The mass of the solutions admitting a power series expansion in $1/r$ at infinity
and preserving the
asymptotic anti-de Sitter geometry
is computed by using
a counterterm subtraction method.
Numerical arguments are presented for the existence of
hairy black hole solutions for a dilaton potential
of the form $V(\phi)=C_1 \exp(2\alpha_1 \phi)+
 C_2 \exp(2\alpha_2 \phi)+C_3$,
 special attention being paid to the case of
${\cal N}=4,~D=4$
gauged supergravity model of Gates and Zwiebach.
\end{abstract}
\section{Introduction}
According to the so called "no-hair" conjecture, a
stationary black hole is uniquely
described in terms of a small set of asymptotically measurable quantities.
This hypothesis was disproved more than
ten years ago,
when several authors presented a counterexample within the framework of SU(2)
Einstein-Yang-Mills (EYM) theory \cite{89}.
Although the new solution was static with vanishing Yang-Mills (YM)
charges, it was different from the Schwarzschild black hole and,
therefore, not characterized by its total mass
(see \cite{Volkov:1999cc} for a comprehensive review of this topic
and an extensive bibliography).

However, much on the literature on hairy black hole solutions
is restricted to the case of an asymptotically flat spacetime.
Since asymptotic flatness is not always an appropriate theoretical
idealisation, and is never satisfied in reality,
it may be important to consider other types of asymptotics,
in particular solutions
with a cosmological constant $\Lambda$.

Asymptotically anti-de Sitter (AAdS) black hole solutions with SU(2) nonabelian fields
have been presented in
\cite{Winstanley:1998sn,Bjoraker:2000qd}.
The  properties of these configurations are strikingly different from
those valid in the asymptotically flat case
(for example there are stable  solutions
in which the gauge field has no nodes; also
solutions exist for continuous intervals of the
parameter space, rather than discrete points).
In an unexpected development, it
has been shown recently that for $\Lambda<0$ even a conformally
coupled scalar field can be painted as hair \cite{Winstanley:2002jt}.

Much of the discussion on AdS hairy black holes
has concerned the case when the matter fields
fall off sufficiently fast such that the conserved charges
can be written as surface integrals involving only the metric and
its derivatives. However,
recently it became clear that the usual AdS-invariant boundary conditions
do not include all AAdS configurations.
Several solutions  involving a minimally coupled self-interacting
scalar field
in the
bulk action and preserving the asymptotic
AdS symmetry group despite the fact that the
standard gravitational mass diverges, have
appeared in the literature \cite{Martinez:2004nb}-\cite{Liu:2004it}.
An exact four-dimensional black hole solution of gravity
with a minimally coupled self-interacting
scalar field has been presented in  \cite{Martinez:2004nb} by
 Martinez, Troncoso and Zanelli (MTZ).
Hairy black hole solutions of ${\cal  N}=8$ gauged supergravity
in four and five dimensions are described in Ref. \cite{Hertog:2004dr}.
Solutions describing a gravitating self-interacting scalar field
whose mass saturates the Breitenlohner-Freedman
bound are discussed in \cite{Henneaux:2004zi}.

In all these cases, the scalar fields drop off so slowly in the asymptotic region,
such that they add a nonzero contribution to the conserved charges.
The mass of these solutions is computed by using an Hamiltonian method,
such that the divergencies
from the gravity and scalar parts cancel out, yielding a finite total charge.

In this paper we consider the case of Einstein gravity coupled to
a dilaton field $\phi$ with a dilaton potential  $V(\phi)$, looking for
solutions satisfying a weakened set of boundary conditions at infinity,
which implies a diverging ADM mass, despite the fact that the spacetime is still AAdS.
To simplify the general picture, we restrict ourselves to the case $D=4$.
Also, since nonabelian fields usually occur together with the dilaton in the bosonic sector
of many gauged supergravity theories, we
include  a SU(2) nonabelian field
in the action (abelian solutions of this theory are discussed in
\cite{Poletti:1994ww}, however for a particular
set of boundary conditions which implies a finite ADM mass).
We suppose that
the dilaton field approaches asymptotically
a constant value $\phi_0$, which corresponds to an extremum of the
potential such that
$ dV/d\phi \big|_{\phi_0}=0$ and $V(\phi_0)<0$. Therefore the configurations
present an effective negative cosmological constant.

Since a negative cosmological constant allows for the existence
of black holes whose horizon has nontrivial topology,
we consider  apart from spherically symmetric solutions,  topological black holes also.
The mass of these solutions is computed by using a counterterm method.

Although we discuss a general case, assuming only the existence of a power series
expansion at infinity, numerical results
are presented mainly for the case of ${\cal  N}=4,~D=4$
gauged supergravity model of Gates and Zwiebach \cite{Gates:1982ct}.

The  paper is structured as follows: in the next Section we explain the model
and derive the basic equations and the asymptotic form of the solutions.
The boundary stress tensor
and the associated conserved charge are computed in Section 3, by using
a counterterm prescription adapted to our case.
The counterterm choice is tested for the MTZ black hole solution.
In Section 4  we present the numerical results,
the case of regular solutions being also briefly discussed.
We conclude with Section 5 where the results are compiled.

\section{General framework and equations of motion}
\subsection{Basic ansatz and field equations}
We start with the following action principle
\begin{eqnarray}
\label{action}
I=\int_{\mathcal{M}} d^4 x \sqrt{-g}
\left(
 \frac{R}{16 \pi G}
-\frac{1}{2} e^{2a \phi}Tr(F_{MN}F^{MN})
-\frac{1}{2}\partial_{M}\phi \partial^{M}\phi-V(\phi)
\right)
-\frac{1}{8 \pi G}\int_{\partial\mathcal{M}} d^3 x\sqrt{-h}K,
\end{eqnarray}
where
$G$ is the gravitational constant, $R$ is the Ricci scalar associated with the
spacetime metric $g_{MN}$.
$F_{MN}=\frac{1}{2} \tau^aF_{MN}^a$ is the gauge field
strength tensor defined as
\begin{equation}
\label{fmn}
F_{MN} =
\partial_M A_N -\partial_N A_M - i g \left[A_M , A_N \right]
\ ,
\end{equation}
where the gauge field is
$ A_{M} = \frac{1}{2g} \tau^a A_M^a,$
$\tau^a$ being the Pauli matrices.
The constant $a$ governs the coupling of $\phi$
to the nonabelian field while $g$ is the gauge coupling constant.
The last term in  (\ref{action}) is the Hawking-Gibbons
surface term \cite{Gibbons:1976ue}, where $K$ is the trace
of the extrinsic curvature for the boundary $\partial\mathcal{M}$
 and $h$ is the induced
metric of the boundary. 

The field equations are obtained by varying the action
(\ref{action})  with respect
to the field variables $g_{MN},A_{M}$ and $\phi$
\begin{eqnarray}
\label{Einstein-eqs}
\nonumber
R_{MN}-\frac{1}{2}g_{MN}R &=&
8\pi G  T_{MN}
\\
\label{dil-eqs}
\nabla^2 \phi-a e^{2a \phi} Tr(F_{MN}F^{MN})
-\frac{\partial V}{\partial \phi}&=&0,
\\
\label{YM-eqs}
\nonumber
\nabla_{M}(e^{2a \phi}F^{MN})-ige^{2a \phi}[A_{M},F^{MN}]&=&0,
\end{eqnarray}
where the energy-momentum tensor is defined by
\begin{eqnarray}
\label{Tij}
T_{MN} =
\partial_{M} \phi \partial_{N} \phi
-\frac{1}{2}g_{MN} \partial_{P}\phi \partial^{P}\phi-
g_{MN}V(\phi)\ +
2e^{2a \phi}{\rm Tr}
    ( F_{MP} F_{NQ} g^{PQ}
   -\frac{1}{4} g_{MN} F_{PQ} F^{PQ}).
\end{eqnarray}
Since for a negative cosmological constant
topological black holes may appear
(whose topology of the event horizon is no longer the two-sphere $S^2$),
we consider a general metric ansatz
\begin{eqnarray}
\label{metric}
ds^{2}=\frac{dr^2}{H(r)}+r^{2}d \Omega_{k}^2-\sigma^2(r)H(r)dt^2,
\end{eqnarray}
where $d \Omega_k^2=d\theta^{2}+f^{2}(\theta) d\varphi^{2}$
is the metric on a two-dimensional surface  of constant curvature $2k$.
The discrete parameter $k$ takes the values $1, 0$ and $-1$
and implies the form of the function $f(\theta)$
\begin{eqnarray}
f(\theta)=\left \{
\begin{array}{ll}
\sin\theta, & {\rm for}\ \ k=1 \\
\theta , & {\rm for}\ \ k=0 \\
\sinh \theta, & {\rm for}\ \ k=-1.
\end{array} \right.
\end{eqnarray}
When $k=1$, the metric takes on the familiar spherically symmetric form,
for $k=-1$
the $(\theta, \varphi)$ sector is a space with constant negative curvature,
while for $k=0$ this is a flat surface
(see $e.g.$ the discussion in \cite{Mann:1997iz}).

Taking into account
the symmetries of the line element (\ref{metric}),
we find the expression of
the purely magnetic YM ansatz \cite{VanderBij:2001ia}
\begin{eqnarray} \label{A}
A=\frac{1}{2g} \left\{
 \omega(r) \tau_1  d \theta
+\left(\frac{d \ln f}{d \theta} \tau_3
+ \omega(r) \tau_2  \right) f d \varphi \right \},
\end{eqnarray}
which gives the YM curvature
\begin{eqnarray}
F=\frac{1}{2g}\left \{
\omega' \tau_1 dr\wedge d\theta +
f \omega' \tau_2 dr\wedge d\varphi +
(w^2-k)f \tau_3 d\theta \wedge d\varphi \right \},
\end{eqnarray}
where a prime denotes a derivative with respect to $r$.

Inserting this ansatz into the
action (\ref{action}), the field equations reduce to
\begin{eqnarray}
\nonumber
\sigma'&=&\frac{8\pi G \sigma}{r}\Big(\frac{e^{2a \phi}}{g^2}\omega'^2
+\frac{1}{2}\phi'^2r^2\Big),
\\
\nonumber
rH'&=& k-H -8 \pi G \Big(
\frac{e^{2a \phi}}{g^2}(\omega'^2 H+\frac{(\omega^2-k)^2}{2r^2})
+\frac{r^2}{2}H\phi'^2 +V(\phi)r^2 \Big),
\\
\label{eqs}
\left(\sigma e^{2a \phi}H\omega' \right)'&=&
\sigma e^{2a \phi} \frac{\omega(\omega^2-k)}{r^2},
\\
\nonumber
\left(Hr^2 \sigma \phi'\right)'&=&2a \sigma  \frac{e^{2a \phi}}{g^2}
\left(
\omega'^2 H+\frac{(\omega^2-k)^2}{2r^2}\right) +
\frac{\partial V}{\partial \phi}r^2 \sigma.
\end{eqnarray}

\subsection{Asymptotic expansion}
We assume that the solution of the above equations
admits at large $r$  a power series expansion of the form
\begin{eqnarray}
\label{asym1}
\phi=\sum_{i=0}^{\infty}\phi_i r^{-i},
~~\omega=\sum_{i=0}^{\infty}\omega_i r^{-i},
~~H= h_0 r^{2} +\sum_{i=2}^{\infty}h_i r^{-i+2},~~
\sigma=\sum_{i=0}^{\infty}\sigma_i r^{-i}.
\end{eqnarray}
From
the lowest order term in equation of $\phi$
we find that $V'_0= dV/d\phi \big|_{\phi_0}=0$
(we shall note $V^{(k)}_0=V^{(k)}(\phi_0)$).
We remark that, by using a suitable redefinition,
we can always set $\phi_0=0$, with no loss of generality.
The effective cosmological constant is
\begin{eqnarray}
\Lambda_{eff}=8\pi G V_0=-3/\ell^2.
\end{eqnarray}
The generic solution has $\lim_{r \to \infty} r^2\phi'  \neq 0$,
which, from the field equations,
implies the following consistency conditions on the dilaton potential
\begin{eqnarray}
\label{cond-pot}
 V^{''}_0=-\frac{2}{\ell^2},~~V'''_0=0.
\end{eqnarray}
Note that the scalar field mass $m^2=V^{''}_0$ is larger than the Breitenlohner-Freedman bound
$m^2=-9/4 \ell^2$.

The assumption (\ref{asym1}) leads to the asymptotic expansion at large $r$
\begin{eqnarray}
\label{expansion1}
H=k+\frac{4 \pi G \phi_1^2}{\ell^2} -\frac{2M_0}{r}+\frac{r^2}{\ell^2}+O(1/r^2),
~~~
\sigma=1-\frac{2\pi G\phi_1^2}{r^2} -\frac{16\pi G\phi_1 \phi_2}{3r^3}+O(1/r^4),
\\
\nonumber
\omega=\omega_0+\frac{\omega_1}{r}
+\frac{\ell^2 \omega_0(\omega_0^2-k)-2a\phi_1 \omega_1}{2r^2}+O(1/r^3),
~~~
\phi=\phi_0+\frac{\phi_1}{r}+\frac{\phi_2}{r^2}+O(1/r^3),
\end{eqnarray}
where $M_0,\omega_0, \omega_1, \omega_2,\phi_0,\phi_1,\phi_2$
are arbitrary constants,
which implies the asymptotic form of the
metric function
\begin{eqnarray}
\label{gtt1}
\nonumber
-g_{tt}= k-\frac{2M+ 32\pi G \phi_1 \phi_2/(3\ell^2)} {r}
+\frac{r^2}{\ell^2}+O(1/r^2),
~~~
g_{rr}=\frac{\ell^2}{r^2}- \frac{\ell^2}{r^4}(4 \pi G \phi_1^2+k\ell^2)
+\frac{2M\ell^4}{r^5}+O(1/r^6).
\end{eqnarray}
For any $\phi_1,~\phi_2$, this set of asymptotics preserve the full AdS symmetry group.
Different from the asymptotically flat case, there are no obvious
restrictions on the value of $\omega_0$.

\section{A computation of mass}
In order to compute quantities like the action and mass
one usually encounters infrared divergences, associated with the infinite volume
of the spacetime manifold.
The traditional approach to this problem is to use the
a background subtraction whose asymptotic geometry
matches that of the solutions.
However, this approach breaks down when there
is no appropriate or obvious background.

In the AdS/CFT inspired counterterm method, this
problem  is solved by adding additional surface terms to the theory action.
These counterterms are built up with
curvature invariants of a boundary $\partial \cal{M}$ (which is sent to
infinity after the integration)
and thus obviously they do not alter the bulk equations of motion.
This yields a well-defined boundary stress tensor and
a finite action and mass of the system.

As found in \cite{Balasubramanian:1999re},
the following counterterms are sufficient to cancel
divergences in four dimensions,
for vacuum solutions with a negative cosmological constant $\Lambda=-3/\ell^2$
\begin{eqnarray}
\label{ct}
I_{\rm ct}^{0}=-\frac{1}{8 \pi G} \int_{\partial {\mathcal M}}d^{3}x\sqrt{-h}\Biggl[
\frac{2}{\ell}+\frac{\ell}{2}\cal{R}
\Bigg]\ ,
\end{eqnarray}
where  ${\cal R}$ is the Ricci scalar
for the boundary metric $h$.

Using these counterterms one can
construct a divergence-free boundary stress tensor $T_{\mu \nu}$
from the total action
$I{=}I_{\rm bulk}{+}I_{\rm surf}{+}I_{\rm ct}^0$ by defining
\begin{eqnarray}
\label{s1}
T_{\mu \nu}&=& \frac{2}{\sqrt{-h}} \frac{\delta I}{ \delta h^{\mu \nu}}
=\frac{1}{8\pi G }(K_{\mu \nu}-Kh_{\mu \nu}-\frac{2}{\ell}h_{\mu \nu}+\ell E_{\mu \nu}),
\end{eqnarray}
where $E_{\mu \nu}$ is the Einstein tensor of the  boundary metric,
$K_{\mu \nu}=-1/2 (\nabla_\mu n_\nu+\nabla_\mu n_\nu)$ is the extrinsic curvature,
$n^M$ being an outward pointing normal vector to the boundary.

If $\xi^{\mu}$ is a Killing vector generating an isometry of the boundary geometry,
there should be an associated conserved charge.
We suppose that the boundary geometry is foliated
by spacelike surfaces $\Sigma$ with metric
$\sigma_{ab}$
\begin{eqnarray}
\label{b-AdS}
h_{\mu \nu}dx^{\mu} dx^{\nu}=-N_{\Sigma}^2dt^2
+\sigma_{ab}(dx^a+N_{\sigma}^a dt) (dx^b+N_{\sigma}^b dt).
\end{eqnarray}
Thus the conserved charge associated with
time translation $\partial /\partial t$ is the mass of spacetime
\begin{eqnarray}
\label{mass}
\mathbf{M}=\int_{ \Sigma}d^{2}x\sqrt{\sigma}N_{\Sigma}\epsilon.
\end{eqnarray}
Here $\epsilon=u^{a}u^{b}T_{a b}$ is the proper energy density
while $u^{a}$ is a timelike unit normal to $\Sigma$.

The presence of the additional matter fields in (\ref{action})
brings the potential danger of having divergent contributions
coming from both the gravitational and matter action \cite{Taylor-Robinson:2000xw}.
For a $1/r^2$ (or faster) decay of the dilaton field
in the asymptotic region, we find that the prescription (\ref{ct})
(with $\ell^2=-3/(8\pi G V_0)$ removes all divergences of the total action,
and implies the usual configurations mass
$
\mathbf{M}=\frac{{\cal  V}}{4 \pi G}M_0,
$
${\cal  V}$ being the area of the ($\theta,~\varphi$) surface (${\cal  V}=4\pi$ for $k=1$).

As expected, a dilaton field which behaves asymptotically as $O(1/r)$
necessarily contributes to the
action and its variations in the asymptotic region.
The counterterms  (\ref{ct}) will not yield in this case
a finite action or mass.
However, similar to the case of  three dimensional
gravity with a minimally coupled scalar field \cite{Gegenberg:2003jr},
it is still  possible to obtain a finite mass by
allowing $I_{ct}$ to depend not only on
the boundary metric $h _{\mu \nu}$, but
also on the scalar field.
This means that the
quasilocal stress-energy tensor (\ref{s1}) also
acquires a contribution coming from the matter field.

Since a polynomial in $\phi$ does not remove the divergencies, we are forced to consider
terms containing the normal derivative of $\phi$.
Following \cite{Gegenberg:2003jr}, we find that by adding a  counterterm of the form
\begin{eqnarray}
\label{Ict}
I_{ct}^{(\phi) }=
\frac{1}{3}
 \int_{\partial M}d^{3}x\sqrt{-\gamma }
\Big(  \phi  {n}^{M}\partial _{M}\phi  +\frac{\ell}{4}m^2\phi^2\Big)
=
\frac{1}{3}
 \int_{\partial M}d^{3}x\sqrt{-\gamma }
\Big(\phi  {n}^{M}\partial _{M}\phi  -\frac{1}{2\ell}\phi^2\Big)
\end{eqnarray}
to the expression (\ref{ct}),
the divergence disappears
\footnote{Note that one can replace the term
$\phi{n}^{M}\partial _{M}\phi$ in (\ref{Ict}) with a suitable
combination of $(n^M \partial _{M}\phi)^2$ and $m^2\phi^2$,
in which case the matter conterterm reads
$I_{ct}^{(\phi) }=
-\frac{\ell}{6}
 \int_{\partial M}d^{3}x\sqrt{-\gamma }
\big(  (n^M \partial _{M}\phi)^2  - m^2\phi^2\big)$,
without changing
the expressions for  boundary stress-tensor, mass and total action.}.

This yields a supplementary contribution to (\ref{s1}),
$T_{ab}^{(\phi)}=1/3g_{ab}(\phi n^M \partial _{M}\phi
- \phi^2/(2\ell))$.
The nonvanishing components of the resulting boundary stress-tensor are
(here we choose $\partial {\mathcal M}$ to be a three surface of fixed $r$,
while $n_{M}=\sqrt{g_{rr}}\delta_{r M}$)
\begin{eqnarray}
\label{BD4}
\nonumber
T_{\theta}^{\theta} =T_{\varphi}^{ \varphi}=
\Big(\frac{2\phi_1 \phi_2}{3\ell}+\frac{M_0}{8 \pi G}\Big)\frac{1}{r^3}
+O\left(\frac{1}{r^4} \right),
~~
T_{t}^t=\Big(-\frac{4\phi_1 \phi_2}{3\ell}-
\frac{2M_0}{8 \pi G}\Big)\frac{1}{r^3}+O\left(\frac{1}{r^4}\right).
\end{eqnarray}
We remark that, to leading order, this stress tensor is traceless
as expected from the AdS/CFT correspondence, since even dimensional
bulk theories are dual to odd dimensional CFTs which have a
vanishing trace anomaly.
Employing the AdS/CFT correspondence,
this result can be interpreted as the expectation value
of the stress tensor in the boundary CFT \cite{Myers:1999qn}.

The mass of these solutions, as computed from (\ref{mass}) is
\begin{eqnarray}
\label{Mct}
\mathbf{M}={\cal  V}\Big(\frac{M_0}{4 \pi G}+\frac{4\phi_1\phi_2}{3\ell^2}\Big).
\end{eqnarray}
This coincides with the mass of the ${\cal N}=8,~D=4$ gauged supergravity solutions
considered in \cite{Hertog:2004dr}, as computed by using an Hamiltonian method
(the asymptotics of those solutions is a particular case of (\ref{expansion1})).
Also, it can be proven that the above counterterm choice
yields a finite Euclidean action.

\subsection{Testing the counterterms with the MTZ exact solution}
Recently Martinez, Troncoso and Zanelli have found an exact black hole solution
of the field equations (\ref{Einstein-eqs}), corresponding to a $F_{\mu \nu}=0$
truncation  of the action (\ref{action})
and a scalar potential \cite{Martinez:2004nb}
\footnote{See also \cite{Nucamendi:1995ex}, \cite{Zloshchastiev:2004ny}
 for another recent examples of scalar hairy black holes and solitons.}
\begin{eqnarray}
\label{potMTZ}
V(\phi )=V_0 \Big(1+2\sinh ^{2} \sqrt{\frac{4\pi G}{3}}\phi\Big)=
-\frac{3}{8\pi G \ell^{2}}\Big(1+2\sinh ^{2} \sqrt{\frac{4\pi G}{3}}\phi\Big).
\end{eqnarray}
The metric and the scalar field are given by
\begin{eqnarray}
\label{MTZ-sol}
ds^{2}&=&\frac{r(r+2G\mu )}{(r+G\mu )^{2}}
\left[
\frac{dr^{2}}
{
  \frac{r^{2}}{\ell^{2}
}-\left( 1+\frac{G\mu }{r}\right) ^{2}
 }+r^{2}(d\theta^2+\sinh^2 \theta d \varphi^2)
 -\left( \frac{r^{2}}{\ell^{2}}%
-\left( 1+\frac{G\mu }{r}\right) ^{2}\right) dt^{2}
\right],
\\
\nonumber
\phi &=&\sqrt{\frac{3}{4\pi G}}\;\mbox{arctanh}\frac{G\mu }{r+G\mu }\;.
\label{scalar}
\end{eqnarray}
\\
The only singularities of the
curvature and the scalar field occur  at $r=0$ and at $r=-2G\mu $.
 These singularities are surrounded by an event horizon located at
\begin{eqnarray}
\nonumber
r_{+}=\frac{\ell}{2}\left( 1+\sqrt{1+4G\mu /\ell}\right),
\end{eqnarray}
while the Hawking temperature is
$T_H= \beta^{-1}=(2 r_+/\ell-1)/(2\pi \ell)$.
A straightforward computation yields the nonvanishing components
of  boundary stress tensor
\begin{eqnarray}
\label{MTZ1}
\nonumber
T_{\theta}^{\theta} =T_{\varphi}^{ \varphi}=
\frac{\mu \ell }{8 \pi r^3}  +O\left(\frac{1}{r^4} \right),
~~
T_{t}^t=-\frac{\mu \ell }{4 \pi r^3} +O\left(\frac{1}{r^4}\right).
\end{eqnarray}
Thus, from (\ref{mass}) we find a total black hole mass
\begin{eqnarray}
\label{Mct1}
\mathbf{M}=\frac{{\cal  V}}{4 \pi}\mu,
\end{eqnarray}
while the total Euclidean action is
\begin{eqnarray}
\label{Mct2}
I=\frac{\beta {\cal  V}}{8 \pi G}(2G\mu+\ell),
\end{eqnarray}
which, from 
\begin{eqnarray}
\label{entropy}
S=\beta \mathbf{M}-I
\end{eqnarray}
gives an entropy which is one quarter of the
event horizon area $A_H$, the first law of thermodynamics being also satisfied.

These results coincide with those found in \cite{Martinez:2004nb}
by using an Hamiltonian formalism.

\section{Numerical solutions}
To solve the field equation (\ref{eqs}), we change to
dimensionless variables by using the rescaling
$r \to (\sqrt{4\pi G}/g) r$,
$\phi \to (1/\sqrt{4\pi G}) \phi$ and
$a \to ( \sqrt{4\pi G}) a$, together with a rescaling of the potential.
\subsection{The form of the dilaton potential}
Given the inherent difficulties involved in studies of these models,
we will restrict ourselves to the case of a potential on the form
\begin{eqnarray}
\label{potential}
V(\phi)=C_1e^{2\alpha_1 \phi}+C_2 e^{2 \alpha_2\phi}+C_3,
\end{eqnarray}
(where we suppose $\alpha_1\neq \alpha_2$).
This type of potential can be obtained when a higher dimensional
theory is compactified
to four dimensions, including various supergravity models
(see \cite{Giddings:2003zw} for a recent discussion of these aspects).

For $-\alpha_1=\alpha_2=1,~C_1=-1/8,~C_2=-\xi^2/8,~C_3=-\xi/2$,
this is the dilaton potential  appearing in the ${\cal  N}=4$ gauged supergravity model
of Gates and Zwiebach.
For $a=1$, the action (\ref{action}) is also a consistent
truncation of the bosonic sector of this model.
Spherically symmetric BPS regular solutions of this theory with the general
asymptotics (\ref{expansion1})
have been constructed recently in \cite{Chamseddine:2004xu}.

Although a Liouville-type potential plus a cosmological constant (obtained for
$C_1$ or $C_2=0$) might be an interesting choice,
one can prove that
such a model  does not possess
solutions with AdS asymptotics.

For a nonzero $1/r$ term in the asymptotic expansion of the dilaton field,
the conditions  (\ref{cond-pot}) impose the following relations
between the potential parameters
\begin{eqnarray}
\label{rel1}
 -\alpha_1=\alpha_2=\alpha,
~~~C_2=C_1 e^{-4\alpha \phi_0},
~~~C_3=2C_1 e^{-2\alpha \phi_0} (3\alpha^2-1).
\end{eqnarray}

By using the scaling properties of the system
$\phi \to \phi+\phi_0,~
r \to r e^{2 a \phi_0}$ we can always set $\phi_0=0$,
resulting in the simple potential
\begin{eqnarray}
\label{pot1}
V(\phi)=C(\sinh ^2 \alpha \phi+\frac{3}{2}\alpha^2),
\end{eqnarray}
where $C=4 C_1 e^{2\phi_0(\alpha-a)}$.

We observe that the potential of the GZ  model can also be written in this form
(for $\xi >0$ $i.e.$ a negative effective cosmological constant),
with $C=-1/2$ and $\alpha=1$.
In this case, following the approach in \cite{Chamseddine:2004xu},
we find the set of Bogomolnyi equations
\begin{eqnarray}
\label{BPS}
\phi'&=&-\frac{r}{2H}F_1F_2,
~~\omega'=e^{-\phi}\frac{\omega r}{2H}F_2,
\\
H&=&\omega^2+\frac{r^2}{2}F_1^2,~~e^{\phi}\omega \sigma=const.,
\\
\label{cons}
\frac{r e^{\phi}F_1}{2H}&=&\frac{H'}{2H}+\frac{\sigma'}{\sigma}-\phi',
\end{eqnarray}
(with $F_1=\cosh \phi +e^{\phi}(k-\omega^2)/r^2$,
$F_2=\sinh \phi +e^{\phi}(k-\omega^2)/r^2)$,
the equation (\ref{cons}) being a differential consequence of the first four equations.
One can verify that these Bogomolnyi equations are compatible with the Eqs. (\ref{eqs}).
Also, there are no black hole solutions
of the above equations.

Numerical arguments for the existence of $k=1$ regular solutions
are presented in \cite{Chamseddine:2004xu}.
The $k=0,-1$ configurations preserving any supersymmetry
present naked singularities.
For example, a $k=0$ solution of the above equations is (with $c$ an arbitrary constant)
\begin{eqnarray}
ds^{2}=\frac{dr^2}{r^2/2+c^2}+r^{2}(d\theta^2+\theta^2 d\varphi^2)-r^2 dt^2,
\\
\nonumber
\phi(r)={\rm arcsinh}\frac{\sqrt{2}c}{r},~~\omega(r)=0,
\end{eqnarray}
presenting a naked singularity at $r=0$.
\subsection{Black hole solutions}
We are interested in black hole solutions having a regular
event horizon at $r=r_h>0$.
The field equations implies
the following behaviour as $r \to r_h$ in terms of three 
parameters ($\phi_h,\sigma_h,\omega_h)$
\begin{eqnarray}
\label{eh}
\nonumber
H(r)&=&\left(\frac{k}{r_h}- e^{2a\phi_h}\frac{(\omega_h^2-k)^2}{r_h^3}
-2V(\phi_h)r_h\right)\left(r-r_h\right)+O(r-r_h)^2,
\\
\nonumber
\sigma(r)&=&\sigma_h
+\frac{2\sigma_h}{r_h}\left(e^{2a\phi_h}\omega'^2(r_h)
+\frac{1}{2}\phi_h'^2r_h^2\right)
(r-r_h)+O(r-r_h)^2,
\\
\omega(r)&=&\omega_h
+\frac{1}{H'(r_h)}\frac{\omega_h(\omega_h^2-k)}{r_h^2}(r-r_h)+O(r-r_h)^2,
\\
\nonumber
\phi(r)&=&\phi_h+\frac{1}{H'(r_h) r_h^2}
\Big (2a e^{2a\phi_h}\frac{(\omega_h^2-k)^2}{2r_h^2}
+\frac{\partial V}{\partial \phi}
\Big|_{\phi_h}r_h^2 \Big)(r-r_h)+O(r-r_h)^2.
\end{eqnarray}
The condition for a regular event horizon is
$H'(r_h)>0$, which places a bound on $\omega_h$
\begin{eqnarray}
\label{mhbound}
 \frac {e^{2a \phi_h}}{r_h ^{2}}\left(\omega_h^{2}-k \right) ^{2}
<k-2V(\phi_h) r_h ^{2},
\end{eqnarray}
and implies the positiveness of the quantity $\omega'(r_h)$.
In the $k=-1$ case, (\ref{mhbound}) implies the existence of
a minimal value of $|V(\phi_h)|$, $i.e.$ for a given $r_h$
\begin{eqnarray}
\nonumber
|V(\phi_h)|>\frac{1}{r_h^2}(1+\frac {e^{2a \phi_h}}{r_h^2}).
\end{eqnarray}
%
%
%
\newpage
\setlength{\unitlength}{1cm}

\begin{picture}(18,7)
\centering
\put(2,0.0){\epsfig{file=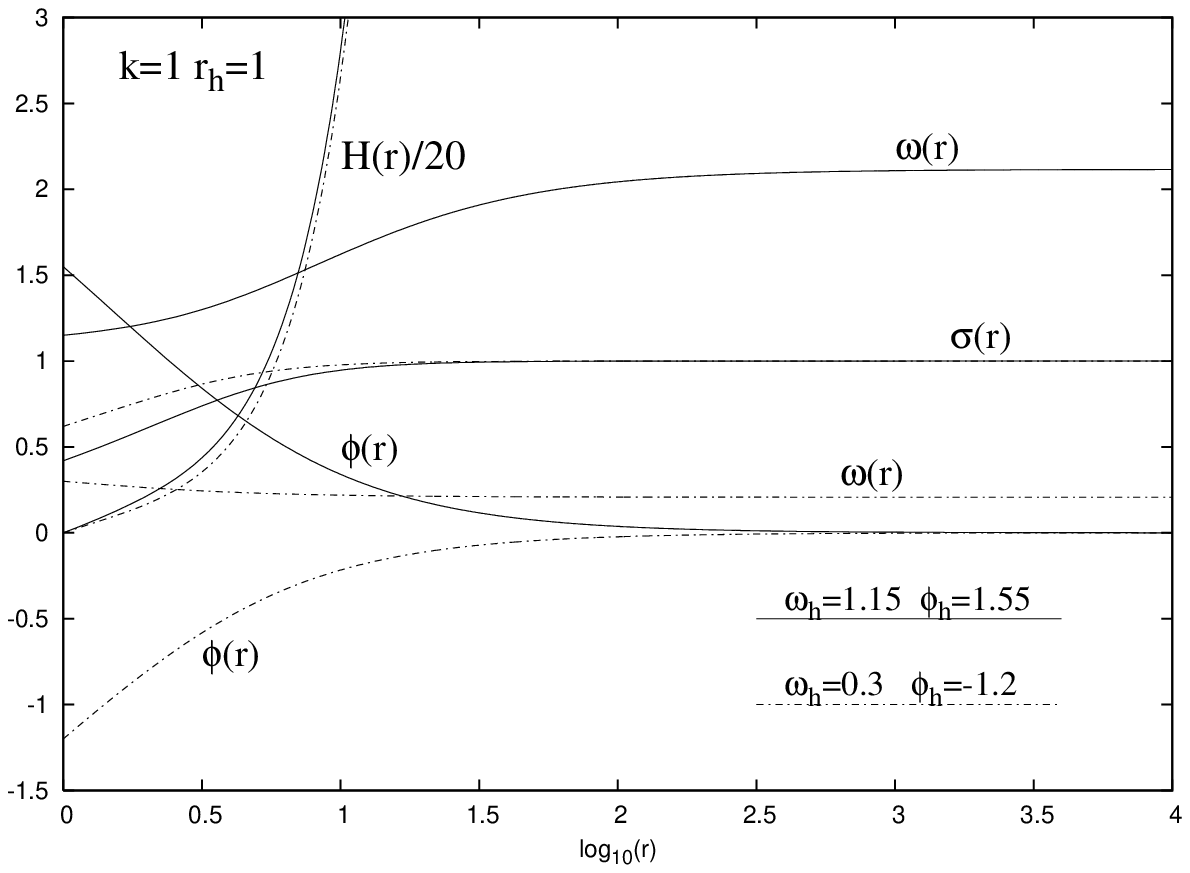,width=12cm}}
\end{picture}
\\
\\
{\small {\bf Figure 1.}
The functions $H(r),~\sigma(r),~\omega(r)$ and $\phi(r)$ are plotted
for two $k=1$ typical black hole solutions}.
\\
\\
%
%
%
By going to the Euclidean section (or by computing the surface gravity)
one find the Hawking temperature
\begin{eqnarray}
\label{TH}
T_H=\frac{1}{\beta}=\frac{\sigma_h H'(r_h)}{4 \pi}.
\end{eqnarray}

Using the initial conditions on the event horizon
(\ref{eh}), the equations were integrated
for $a=\alpha=1$, several values of $C$, and a large set of $\omega_h,~\phi_h$.
Also, although we present here only the case $r_h=1$, similar
solutions seem to exist for any value of $r_h$.
Since the equations (\ref{eqs})  are invariant under the
transformation $\omega \rightarrow - \omega $, only values of
$\omega _{h}\geq 0$ are considered. Nontrivial black hole
solutions with scalar hair exist also in the absence of a gauge field
\footnote{For example, for $k=-1$ and the dilaton potential (\ref{potMTZ})
we have found a family of solutions in terms of $\phi_h$, 
the exact MTZ solution (\ref{MTZ-sol}) being a particular case.}.

The basic properties of these solutions are very similar to
the known EYM-$\Lambda$ black holes \cite{Winstanley:1998sn,Bjoraker:2000qd,VanderBij:2001ia}.
For any $\phi_h$ and $C<0$, solutions appear for continuous intervals
of $\omega_h$ (for $C=-1/2$, we find always only one interval).
For small $|V_0|$, these intervals are separated by intervals
on which there are no solutions with the  asymptotics (\ref{expansion1}).
As $\omega_h$ approaches some critical value $\omega_h^c$,
the metric function $\sigma(r)$ approaches a zero value
on the event horizon.
The value of $\omega_h^c$ increases as $|V_0|$ increases.
Also, there are  black hole solutions for which $\omega_0>1$ although $\omega_h<1$.
Typical spherically symmetric solutions of the GZ model are presented in Figure 1.

For $k=0,-1$, in contrast to the spherically symmetric case,
we find only nodeless solutions, for all values of the parameters.
This can be analytically proven by
integrating the equation for $\omega$,
$(\sigma e^{2a \phi}H\omega' )'=
\sigma e^{2a \phi} \omega(\omega^2-k)/r^2$ between $r_h$ and $r$;
thus obtaining $\omega'>0$ for every $r>r_h$.
For $k=1$ and $|V_0|$ sufficiently large (i.e. $|V_0|>0.01$),
there also exist solutions
for which the gauge function $\omega$ has no nodes.
Increasing the value of $|V_0|$, the ratio
$\omega_0/\omega_h$ remains close to one for most of the $\omega_h$ interval,
and we find nodeless solutions only.

The properties of typical solutions are presented in Figure 2 for $k=1,-1$
and a GZ potential (a similar picture is found for $k=0$).
We observe that the generic solutions we find do
not have the usual AdS asymptotics since
$\phi_1=-\lim_{r \to \infty} r^2\phi' \neq 0$.
However, for any $(V_0,~\omega_h)$, solutions with $\phi_1=0$
exist for a discrete set of $\phi_h$.

Both the parameter $M_0$ and the total mass $\mathbf{M}$ of some $k=-1$ solutions
is negative, a common situation in the topological black hole physics.
For $k=1,~0$ we find $\mathbf{M}>0$ for all boundary conditions we have considered.
\newpage
\setlength{\unitlength}{1cm}

\begin{picture}(18,7)
\centering
\put(2,0.0){\epsfig{file=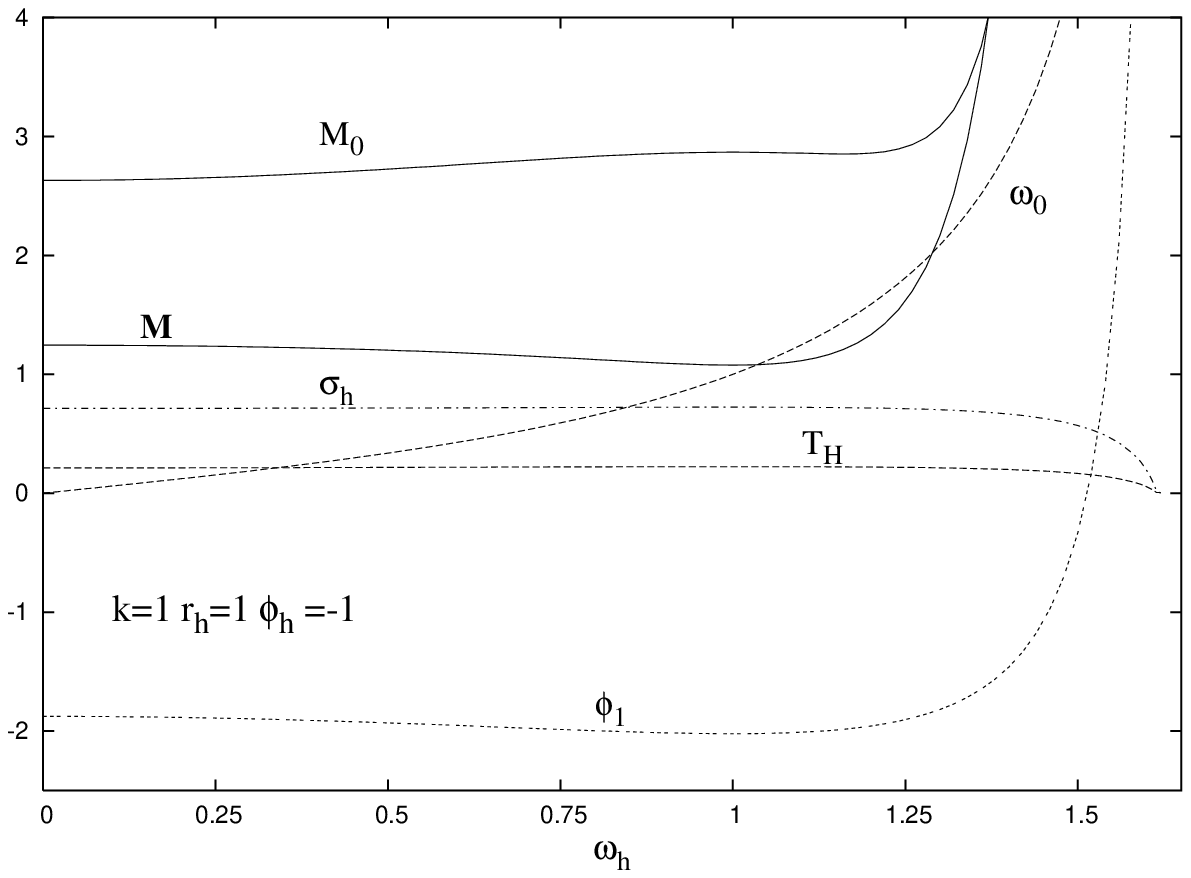,width=12cm}}
\end{picture}
\begin{picture}(19,8.5)
\centering
\put(2.6,0.0){\epsfig{file=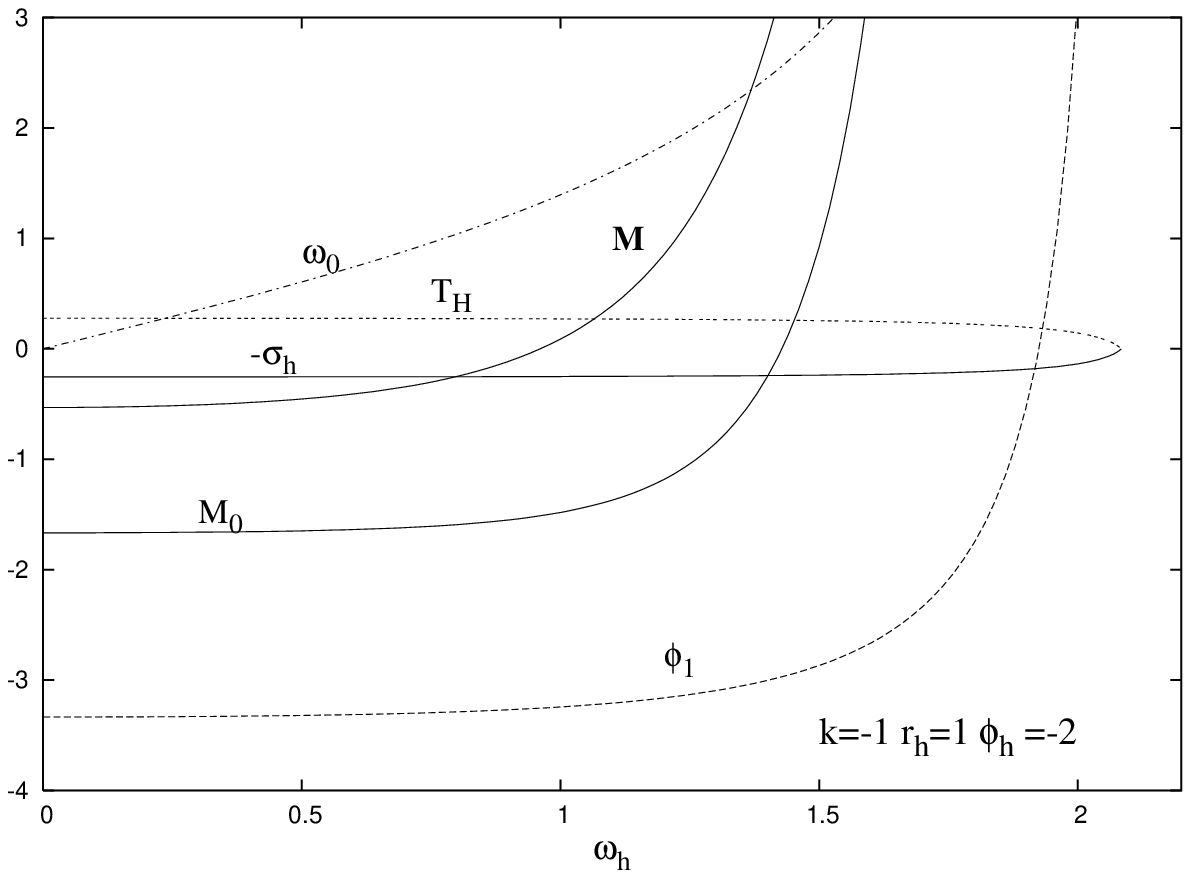,width=12cm}}
\end{picture}
\\
\\
{\small {\bf Figure 2.}
The asymptotic
parameters $\omega_0,~M_0,~\phi_1$, the total mass
$\mathbf{M}$, the Hawking temperature $T_H$  and the value $\sigma_h$ of the metric function
function $\sigma(r)$ at the event horizon are shown as a function of
$\omega(r_h)$ for spherically symmetric and
$k=-1$ topological black hole solutions of the GZ model.
}
\vspace{0.5cm}
\\
It can be proven that the entropy of these black holes is one quarter of
the event horizon area, as expected. By integrating the Killing identity
$\nabla^a\nabla_b K_a=R_{bc}K^c,$
for the Killing field $K^a=\delta^a_t$, together with the Einstein equation
(here we have set $4 \pi G=1$)
\begin{eqnarray}
\label{Rtt}
\frac{1}{2}R_t^t=
\frac{R}{4}
-\frac{1}{2} e^{2a \phi}Tr(F_{\mu \nu}F^{\mu \nu})
-\frac{1}{2}\partial_{\mu}\phi \partial^{\mu}\phi-V(\phi),
\end{eqnarray}
it is possible to isolate the bulk action contribution at infinity and on the event horizon.
The counterterms discussed  in Section 3 regularize the infrared divergencies, such that
the contribution  from the asymptotic region to the total action is found to be
${\cal V}\beta (M_0+ 4\phi_1 \phi_2/ (3\ell^2))$, the relation $S=A_H/4$
resulting straightforwardly from
(\ref{entropy}), (\ref{TH}).
\newpage
\setlength{\unitlength}{1cm}

\begin{picture}(18,7)
\centering
\put(2,0.0){\epsfig{file=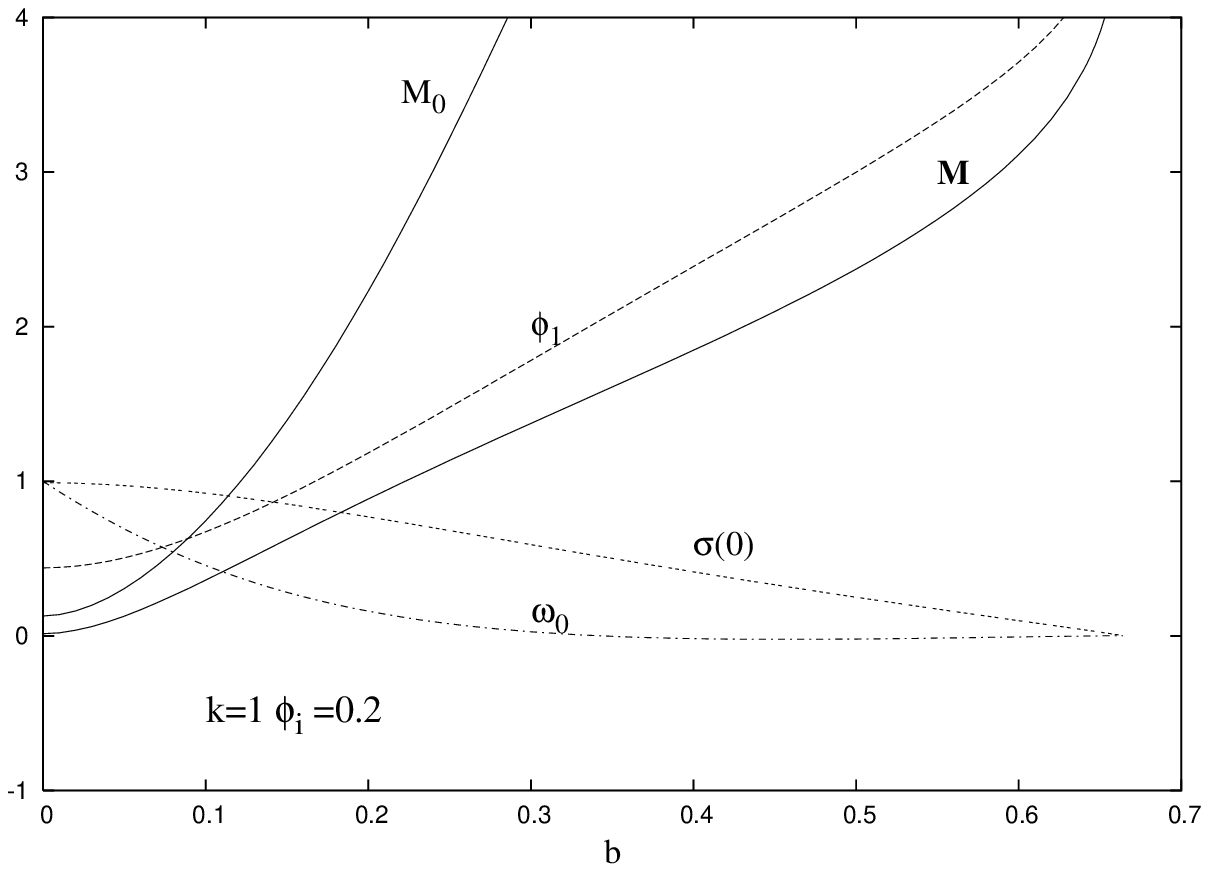,width=12cm}}
\end{picture}
\\
\\
{\small {\bf Figure 3.}
The asymptotic parameters $\omega_0,~M_0,~\phi_1$, the total mass
$\mathbf{M}$ and the value $\sigma_0$ of the metric function
function $\sigma(r)$ at the origin are shown as a function of
$b$ for typical spherically symmetric regular solutions of the GZ model}.
\vspace{0.3cm}
%
%
\subsection{Regular solutions}
The existence of regular counterparts of the black hole solutions
is possible in the spherically symmetric case only.
For $k\neq 1$, a direct inspection of the system (\ref{eqs}) reveals the absence 
of solutions with a regular origin.
In this case, it is not possible to take a consistent set 
of boundary conditions at the origin 
 without introducing a curvature singularity at $r=0$.
This fact has to be attributed to the particular form of the potential term
$V_{YM}=e^{2a \phi} (\omega^2-k)^2/(2r^2)$, which is fourth order in the YM function
$\omega$, in the reduced lagrangean of the system,.
We  observe the similarity with the EYM system with $\Lambda<0$, where the absence 
of $k \neq 1$ regular configurations has been noticed in \cite{VanderBij:2001ia}.


For completeness  we present here 
the basic properties of the $k=1$ regular configurations.

The behaviour of regular solutions near the origin is
\begin{eqnarray}
\label{origin}
H(r)=1-\left(4b^2e^{2a\phi_i}+\frac{2}{3}V(\phi_i)\right)r^2+O(r^3),
~~~
\sigma(r)=\sigma_0 (1+4 e^{2a\phi_i}b^2)r^2+O(r^3),
\\
\nonumber
\omega(r)=1-b r^2+O(r^4),
~~~
\phi(r)=\phi_i+ (2a e^{2a\phi_i}b^2+\frac{1}{6}\frac{d V}{d \phi}\Big|_{\phi_i})r^2+O(r^3),
\end{eqnarray}
where $b,~\sigma_0$ and $\phi_i$ are arbitrary parameters.
We note that for BPS solutions of the GZ model,
there are only two independent parameters since in this case
$b=1/6e^{-\phi_i}\sinh \phi_i$.

The overall picture we find
is similar to the one described in \cite{Bjoraker:2000qd} for the EYM-$\Lambda$ system
(here we consider again the case of the GZ model).
By varying the parameters $b,\phi_0$, a continuum of monopole solutions is obtained.
As $b$ increases, the value at the origin of the  metric function $\sigma(r)$ decreases and,
for some finite values of $b$, a singularity appears.
The total mass of the solutions as given by (\ref{Mct}) is  an increasing function of $b$.
For any value of $b$ it seems to be
always possible to find a initial value of the scalar field such that $\phi_1=0$.

As seen in Figure 3, the solution with $b=0$ ($i.e.$ $\omega(r)\equiv 1$) is not the vacuum
AdS spacetime.
Thus, rather unexpectedly,
for these asymptotics there are regular solutions even without a nonabelian field.
For the GZ model, we find a continuum of (nonsupersymmetric-) scalar solitons,
as a function
of the value of the scalar field at the origin $\phi_i$.
A similar property has been noticed recently for a truncation of ${\cal  N}=8,~D=4$
gauged supergravity \cite{Hertog:2004jx}.

The  expansion  as $r \to \infty$ is valid also for these regular solutions.
For the GZ model, the coefficients of BPS solutions in the asymptotics (\ref{expansion1})
satisfy the relations $\phi_2=k-\omega_0^2,~M=\phi_1(\omega_0^2-k)$,
$\omega_1=\omega_0 \phi_1$ 
(we recall that for $k \neq 1$, these describe  configurations with a naked singularity at $r=0$).
We note also the intriguing
expression for the total mass of BPS configurations
\begin{eqnarray}
\label{mass-bps}
\mathbf{M}_{BPS}=\frac{1}{3}Q_D Q_M,
\end{eqnarray}
with $Q_D=-\phi_1,~Q_M=k-\omega_0^2$.

\section{Conclusions.}
 If the gravitating matter
fields do not fall off sufficiently fast at infinity,
 the asymptotic behaviour of the metric can be different
from that in pure gravity.
By relaxing the standard asymptotic conditions for AAdS solutions, it is possible
to preserve the original symmetries at infinity,
while the conserved charges are modified by including matter field terms.

The aim of this letter was to consider this situation for
a theory including, apart from an SU(2) nonabelian field,
 a dilaton field with a nontrivial
potential, playing the role of a cosmological term.
We have found that the addition of the scalar potential
greatly increases the wealth of
possible solutions, preserving at the same time all
features familiar from the EYM-$\Lambda$ case.
For solutions admitting an asymptotic power series expansion,
we have proposed a counterterm choice which gives  finite mass and Euclidean action.
The results we have found are in agreement with those obtained via the Hamiltonian method.

Numerical solutions have been presented mainly for the  case
of a consistent truncation of ${\cal  N}=4,~D=4$ GZ gauged supergravity model.
Both regular and black hole solutions have been presented, the solutions with a finite ADM
mass constituing a discrete set.
By using the relations in \cite{Cvetic:1999au},
we can uplift these configurations to $D=11$, which may suggest
a holographic interpretation for them.
As observed in \cite{Hertog:2004dr},
according to the general AdS/CFT correspondence, there should be a dual
CFT corresponding to each choice of boundary conditions.

Similar to the EYM-$\Lambda$ case, we expect some of
the  solutions with no nodes in the nonabelian
magnetic field to be stable against linear perturbations.

Also, it is possible to relax the asymptotic asumptions (\ref{asym1}), 
allowing a generic noninteger decay at infinity.
In this case, apart from $ V_0=-3/(8\pi G\ell^2),~~V'_0=0$
there are no other restrictions on the dilaton potential and we find
the  asymptotic behaviour
(with $m^2=V''_0<0$)
\begin{eqnarray}
\label{gen-as}
H&=&k-\frac{2M_0}{r}+\frac{r^2}{\ell^2}+
f(m,\ell,\phi_1)
r^{\lambda_+ - \lambda_--1}+\dots,~~
\sigma=\exp(2\pi G \lambda_- \phi_1^2/r^{2\lambda_-})+\dots,
\\
\nonumber
\phi&=&\frac{\phi_1}{r^{\lambda_-}}+\frac{\phi_2}{r^{\lambda_+}}+\dots,~~~
\omega=\omega_0+\frac{\omega_1}{r}+\dots,
\end{eqnarray}
where
\begin{eqnarray}
\label{gen-as2}
\lambda_{\pm}=\frac{3\pm \sqrt{9+4m^2 \ell^2}}{2},~~
f(m,\ell,\phi_1)=-4 \pi G \phi_1^2 \frac{m^2+\lambda_-^2/\ell^2}{\lambda_+-\lambda_-}.
\end{eqnarray}
Both regular and black hole solutions with these asymptotics are likely to exist.
It can be proven that the counterterm choice (\ref{Ict}) gives a finite
mass and action also in this case, yielding very similar results
to those derived in this paper for
$\lambda_-=1,~\lambda_+=2$.

It would be interesting to generalize
these solutions to higher dimensions
 and to find the general matter counterterms expression.
\\
\\
{\bf Acknowledgement}
\\
This work was carried out in the framework of Enterprise--Ireland
Basic Science Research Project SC/2003/390.



\end{document}